\documentclass[aps,pra,reprint,superscriptaddress,showpacs]{revtex4-1}

\usepackage{amsmath}
\usepackage{graphicx}
\usepackage{natbib}

\begin{document}

\title{Transition to lasing induced by resonant absorption}

\author{I. A. Nechepurenko}
\affiliation{Moscow Institute of Physics and Technology, 9 Institutskiy per., Dolgoprudny 141700, Russia}
\affiliation{All-Russia Research Institute of Automatics, 22 Sushchevskaya, Moscow 127055, Russia}

\author{D. G. Baranov}
\email[]{denis.baranov@phystech.edu}
\affiliation{Moscow Institute of Physics and Technology, 9 Institutskiy per., Dolgoprudny 141700, Russia}
\affiliation{All-Russia Research Institute of Automatics, 22 Sushchevskaya, Moscow 127055, Russia}

\author{A. V. Dorofeenko}
\affiliation{Moscow Institute of Physics and Technology, 9 Institutskiy per., Dolgoprudny 141700, Russia}
\affiliation{All-Russia Research Institute of Automatics, 22 Sushchevskaya, Moscow 127055, Russia}
\affiliation{Institute for Theoretical and Applied Electromagnetics, 13 Izhorskaya, Moscow 125412, Russia}

\date{\today}

\begin{abstract}
We theoretically demonstrate that increase of absorption with constant gain in laser systems can lead to onset of laser generation. This counter-intuitive absorption induced lasing (AIL) is explained by emergence of additional lasing modes created by an introduction of an absorbing medium with narrow linewidth. We show that this effect is universal and can be encountered in simple Fabry-Perot-like systems and doped spherical dielectric nanoresonators. The predicted behavior is robust against frequency detuning between the gain and absorbing medium.
\end{abstract}

\pacs{42.25.Bs, 42.55.Ah, 42.60.Da}

\maketitle

\section{Introduction}
It is generally assumed that an increase of loss in a laser system leads to decrease of lasing intensity and, eventually, makes laser turn off. On the other hand, it is also widely accepted that increase of gain correspondingly increases laser radiation intensity. This na\"{\i}ve picture stems from the primitive single-mode rate equations, which are often employed for description of laser operation \cite{Haken,Milonni,Sargent}. Nevertheless, behavior of lasing modes can be much more complicated. Particularly, it has been reported recently that increase of pump intensity in a laser system with non-uniform gain distribution makes the laser turn off \cite{Liertzer2012}. This striking behavior of a laser is induced by exceptional points in the spectra of the non-Hermitian Hamiltonian describing response of the laser system at which two different eigenvectors of the Hamiltonian coalesce. Importantly, such unusual effect was found only in a laser cavity with spatially varying gain profile. The behavior of exceptional points in systems of coupled lasers was investigated in details within the framework of the developed model in Ref. \cite{Ganainy}. Laser turn-off was experimentally demonstrated in a configuration involving a pair of coupled microdisk lasers \cite{Brandstetter} and in a pair of active RLC-circuits \cite{Chitsazi}. Finally, in a recent experimental work the counterpart of lasing shutdown was observed: authors reported that adding spatially distributed loss in a pair of coupled wispering gallery mode resonators may increase lasing intensity \cite{Peng}. Again, this behavior relies on presence of exceptional points in the spectrum of model Hamiltonian.

Usually, laser modes are formed from the quasistationary states, or scattering resonances, of the passive cavity \cite{SALT}. These states are called quasistationary since their eigenfrequencies are, basically, complex-valued what indicates their exponential decay in the absence of applied gain. Only recently it was shown that lasers based on low $Q$ cavities with spatially varying gain-loss profile posses novel modes which can not be attributed to the modes of the passive cavity \cite{Ge}. These modes are related to transmission resonances of a slab appearing with non-uniform distribution of gain and are absent when no gain is applied to the cavity. There are other examples of complicated laser behavior induced by interplay of gain and absorption. For example, in Ref. \cite{Liu2013} intrinsic self-absorption in semiconductors was utilized in order to tailor lasing wavelength of the nanowire lasers.

In this paper, we show that increase of loss in a laser system brought by a dispersive absorbing medium can lead to emergence of lasing. In contrast to experiment by Peng et al. \cite{Peng}, spatial inhomogeniuty of loss or gain distribution is not required in our configuration. Instead, the predicted phenomenon of absorption induced lasing (AIL) is related not to geomtery of the system, but to frequency dispersion of the resonator medium. Within the framework of scattering matrix formalism, we encounter this effect in a simple toy system, consisting of a planar slab made of a uniform mixed loss-gain medium. Similarly to unconventional "surface" laser modes \cite{Ge}, absorbing medium induces new system of $S-$matrix poles corresponding to quasistationary modes. With increase of loss, these additional modes evolve and begin to lase. We also predict AIL in an alternative spherical laser resonator which is free of certain unwanted effects that may occur in planar laser cavities.

\section{Absorption induced lasing in planar cavities} 
Although laser generation is essentially nonlinear process, below threshold laser can be treated as a linear system with negative imaginary part of the gain medium which indicates population inversion of quantum emitters created by external pump \cite{Haken}.
In the linear approximation the lasing threshold is associated with the first pole of the $S-$matrix eigenvalue having real-valued frequency \cite{SALT}.
This pole represents a solution to the Maxwell's equations without incident field.
In absence of gain all poles are located in the lower half-plane of the complex frequency plane, and all corresponding solutiuons exponentially decay in time.
At a certain level of gain pole crosses axis of real-valued frequencies giving rise to self-sustained undamped oscillations.
With further increase of gain, pole will move to the upper half-plane, $\operatorname{Im} \omega>0$, indicating birth of a laser instability with the corresponding time dependence of electromagnetic fields proportional to $\exp \left( {\operatorname{Im} \omega t} \right)$.
Eventually laser oscillations reach steady-state regime and this exponential growth is supressed by saturation of the gain medium.
Though any information about amplitude of laser oscillations can not be obtained from the linear theory, it easily allows one to retrieve the region of laser generation as the area of the system parameters when at least one $S-$matrix singularity is located in the upper frequency half-plane. 

The toy system we study is a slab of \emph{uniform} medium which simultaneously contains linear loss and gain. Gain medium is described by permittivity \cite{Solimeno} which is a sum of two Lorentzian contributions representing absorption and gain:

\begin{equation}
\varepsilon \left( k \right) = {\varepsilon _0} + \frac{  { 2f_A{k_A}{\gamma _A}}}
{{k_A^2 - {k^2} - 2ik{\gamma _A}}} - \frac{  { 2f_G {k_G}{\gamma _G}}}
{{k_G^2 - {k^2} - 2ik{\gamma _G}}},
\label{eq1}
\end{equation}
where $\varepsilon _0$ is the background permittivity of "cold" umpumped medium, ${k _A}$ and $k_G$ are the central frequencies of the absorption and emission (gain) lines, respectively, $\gamma_A$ and $\gamma_G$ are their linewidths and $f_A$ and $f_G$ are corresponding oscillator strengths for the absorption and emission lines. Speed of light for simplicity is assumed to be $c=1$. For a uniform slab of thickness $L$ the two eigenvalues of the scattering matrix for normal incidence read:

\begin{equation}
{s _ \pm } = T \pm R = \frac{{\left( {\exp \left( {iknL} \right) \pm r} \right)\left( {1 + r\exp \left( {iknL} \right)} \right)}}{{1 - {r^2}\exp \left( {2iknL} \right)}},
\label{eq2}
\end{equation}
with $n=\sqrt{\varepsilon}$ being refractive index of the slab medium and $r = {{\left( {1 - n} \right)} \mathord{\left/
 {\vphantom {{\left( {1 - n} \right)} {\left( {1 + n} \right)}}} \right.
 \kern-\nulldelimiterspace} {\left( {1 + n} \right)}}$ Fresnel reflection coefficient from the semi-infinite space. Poles of both eigenvalues $s_ \pm$ coincide and occur when the denominator vanishes:

\begin{equation}1 - {r^2}\exp \left( {2inkL} \right) = 0.
\label{eq3}\end{equation}
Note that if $r$ has a pole, than the whole expression~(\ref{eq2}) is finite. If Eq.~(\ref{eq3}) holds for a real-valued frequency $k$, then it turns into the condition of lasing threshold. This condition can be rewritten in the form of two separate real-valued equations:

\begin{subequations}
\begin{eqnarray}
\operatorname{Re} \left[ {2in{k}L + 2{\text{Ln }}r} \right] = 0,\label{eq4a}
 \\ 
\operatorname{Im} \left[ {2in{k}L + 2{\text{Ln }}r} \right] = 2\pi m, \label{eq4b}
\end{eqnarray}
\end{subequations}
where ${\text{Ln }}r = \ln \left| r \right| + i{\text{Arg }}r$ is the complex logarithm. First of these two equations represents the amplitude condition for lasing. When Eq.~(\ref{eq4a}) is fulfilled, gain compensates the losses due to absorption inside the resonator and radiation into the free space. The second of these two expressions, Eq.~(\ref{eq4b}), is the phase condition of lasing. If condition~(\ref{eq4b}) holds a plane wave travelling inside the resonator acquires phase $2\pi m$ on a length of the slab, as is necessary for formation of a laser mode.

\begin{figure} 
\includegraphics[width=0.4\textwidth]{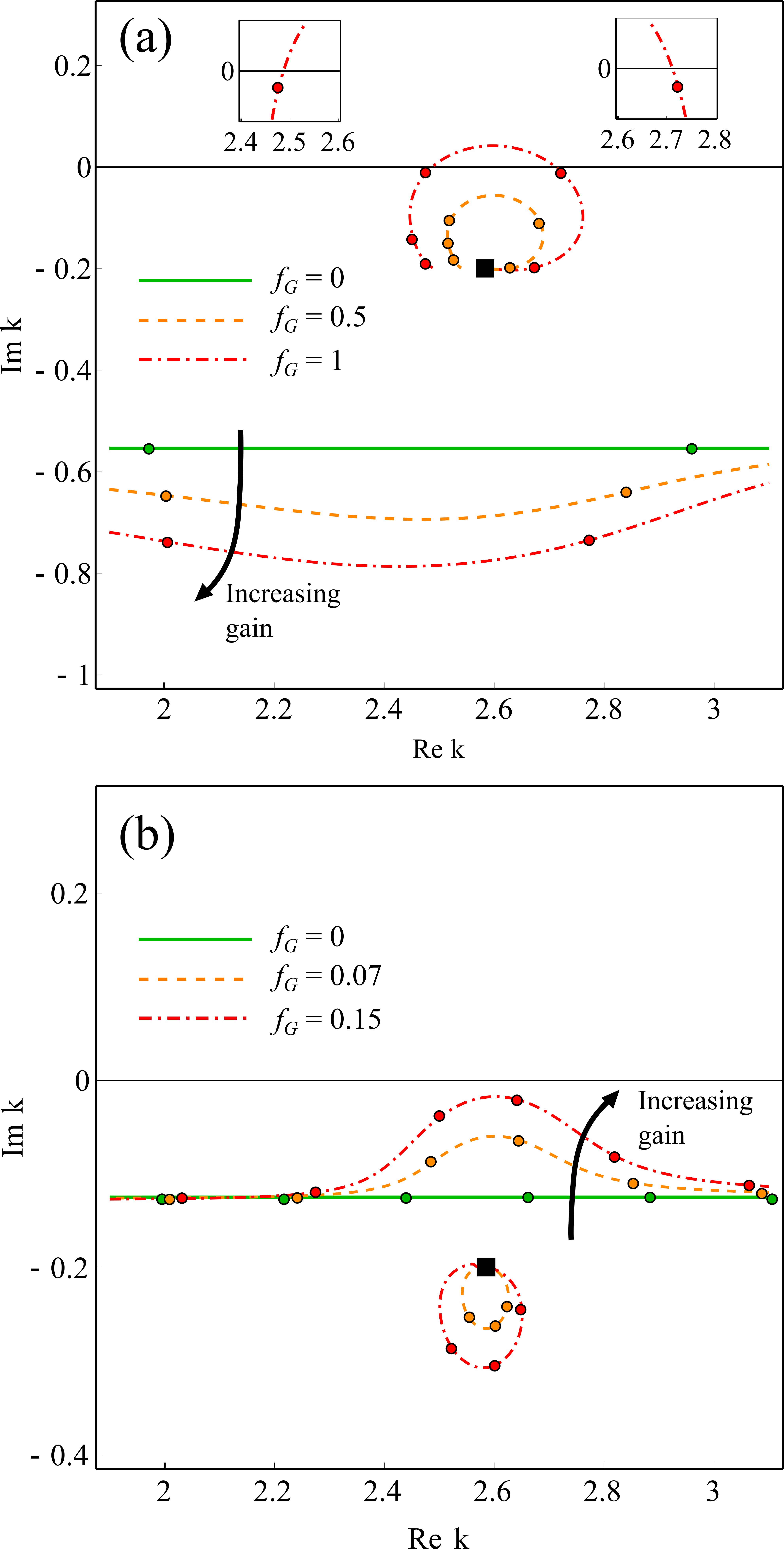}
\caption{\label{fig1}(a) Evolution of $S-$matrix poles for the case of a low $Q$ resonator formed by a slab of length $L = 2.25 \mu m$ with permittivity of gain medium in the absence of pumping $\varepsilon_0=2$, emission frequency of gain medium $k_G=2.6 \mu m^{-1}$ and linewidth of gain medium ${\gamma _G} = 0.2 \mu m^{ - 1}$ in the range of gain parameter $0 \le {f_G} \le 1$. Curves denote the solution of the amplitude condition~(\ref{eq4a}). Solid curve: passive system ($f_G=0$), dashed curve: $f_G=0.5$, dot-dashed curve: $f_G=1$. The box shows the point of poles condensation, located exactly at $k=k_G - i\gamma_G$. Insets show zoomed area around the two poles lying close to the real axis. (b) The same as (a) for the case of a high $Q$ resonator of length $L=10 \mu m$.}
\end{figure}

Fig.~\ref{fig1}(a) shows evolution of the scattering matrix poles occuring with increase of gain in the case of a low $Q$ cavity. Low $Q-$factor of the cavity is forced by choosing  short optical length of the slab $nL \lesssim \lambda$. Curves depicted on the graph in Fig.~\ref{fig1}(a) are obtained by solving Eq.~(\ref{eq4a}). All poles of the passive resonator are located in the lower half-plane indicating dissipative response of the planar system. When gain with central frequency $k_G$ and linewidth $\gamma_G$ is added to the laser, poles of the passive cavity change their position. But, more importantly, a new system of poles located on a circle around complex frequency $k=k_G-i\gamma_G$ emerges.

In fact, the $S-$matrix singularities associated with a dispersive Lorenzian medium (either gain or absorbing) represent damped oscillations of two-level emitters constituing the medium. The gain (absorbing) term in the permittivity expression~(\ref{eq1}) has a pole in the complex frequency plane exactly at $k=k_G-i\gamma_G$ ($k=k_A-i\gamma_A$). This pole of permittivity (not of the $S-$matrix!) corresponds to bulk oscillations of classsical oscillators modeling two-level systems inside unbound Lorenzian medium. When medium is bound to a certain geometry, this bulk oscillation mode couples to the resonator modes defined by its geomtery and splits into a sequence of the $S-$matrix poles. The complex eigenfrequency of the Lorenzian medium then becomes the point of the $S-$matrix poles condensation.

In the case of a hiqh $Q$ resonator, illustrated in Fig.~\ref{fig1}(b), increase of gain shifts poles of the passive cavity towards the real axis, and the mode of the passive cavity starts to lase at threshold. Going back to Fig.~\ref{fig1}(a), we note that in the opposite case of a low $Q$ cavity, which is the subject of current paper, situation is different: poles associated with the emission line of gain medium emerge \emph{closer} to the real axis than poles of the passive resonator, and increase of gain make poles of the passive cavity move away from the real axis. These poles, just as those of the passive cavity, can cross real axis and form lasing modes. One observes that for gain $f_G=0.5$ poles of the scattering matrix are still located in the lower half-plane. Further, when gain reaches value $f_G=1$ there is no laser generation yet, but, remarkably, part of the amplitude curve~(\ref{eq4a}) (dot-dashed curve) apperas in the upper frequency half-plane. In fact, there is enough power in the system to support lasing, nevertheless, lasing modes do not arise in the corresponding frequency range because the phase condition~(\ref{eq4b}) is not fulfilled. At this moment, the system is in the \emph{overpumped} state.

\begin{figure} 
\includegraphics[width=0.45\textwidth]{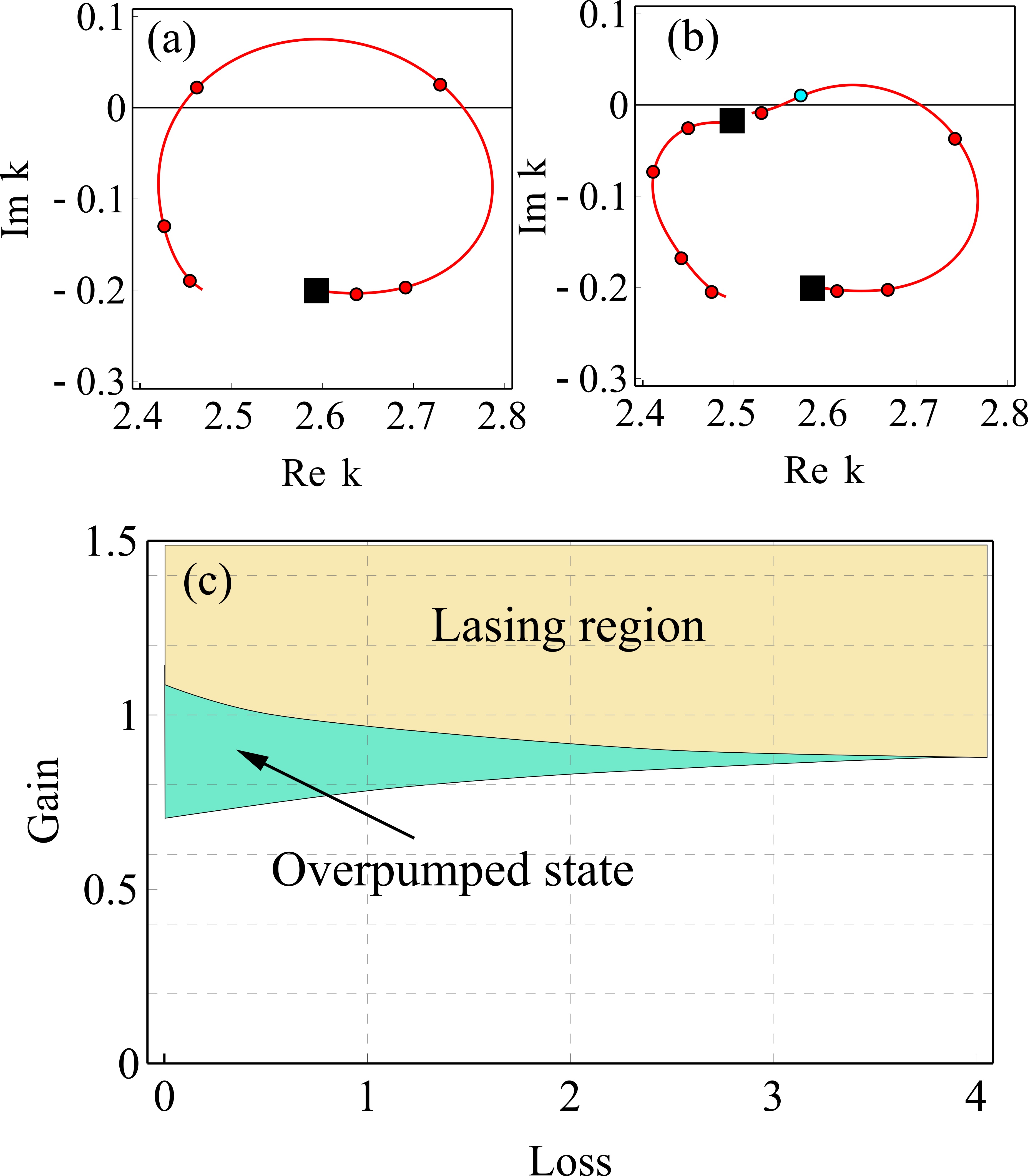}
\caption{\label{fig2}Two scenarios of the $S-$matrix poles evolution after reaching the overpumped state. (a) Position of poles of the $L=2.25 \mu m$ cavity on the complex frequency plane for $f_G=1.2$. Curve denotes the amplitude condition (\ref{eq4a}). (b) The same when absorption with central frequency ${k_A} = 2.5 \mu m^{ - 1}$ and linewidth ${\gamma _A} = {\gamma _G}/10$ is added to the system. Gain and absorption values are set to $f_G=1$ and $f_A=2$. Pole denoted by cyan color enters the upper half-plane. (c) Overpumped and lasing regions in the gain/loss parameters space.}
\end{figure}

After reaching the overpumped state of the laser at $f_G=1$, two different scenarios shown in Fig.~\ref{fig2} are possible. If gain increases even more up to $f_G=1.2$, Fig.~\ref{fig2}(a), the $S-$matrix poles will eventually cross the real axis and lasing will start. However, there is another way to attain lasing \emph{without further increase of gain}. To do so, we introduce an absorbing medium having a narrow resonance (${\gamma _A} \ll {\gamma _G}$) with central frequency ${k_A}$ slightly detuned from the emission line of gain medium. Analogously to addition of gain, introducing of absorbing Lorenzian medium leads to the emergence of additional set of poles (quasistationary eigenstates). For a very narrow absorption line those poles emerge close to the real axis near the poles condensation point $k = {k_A} - i{\gamma _A}$. Fig.~\ref{fig2}(b) shows the position of poles for amount of absorption corresponding to $f_A=2$. At least one scattering pole moves to the upper half-plane indicating laser generation. Here, the amplitude curve is still in the upper half-plane, but it is the absorbing medium which allows to fulfill the phase condition. Although losses exceed gain within the narrow absorption line, i.e. $\operatorname{Im} \varepsilon  > 0$, position of poles is changed due to additional frequency dispersion of the slab medium permittivity $\varepsilon \left( k \right)$ brought by absorption, and certain scattering poles may enter the lasing region $\operatorname{Im} k > 0$. Fig.~\ref{fig2}(c) presents the regions of lasing and overpumped system in the space of loss/gain parameters. Negative incline of the curve dividing the lasing and overpumped regions illustrates the fact that one can attain lasing with increase of absorption. Another interpretation of this feature of Fig.~\ref{fig2}(c) is that lasing threshold decreases with increase of absorption.

\begin{figure} 
\includegraphics[width=0.4\textwidth]{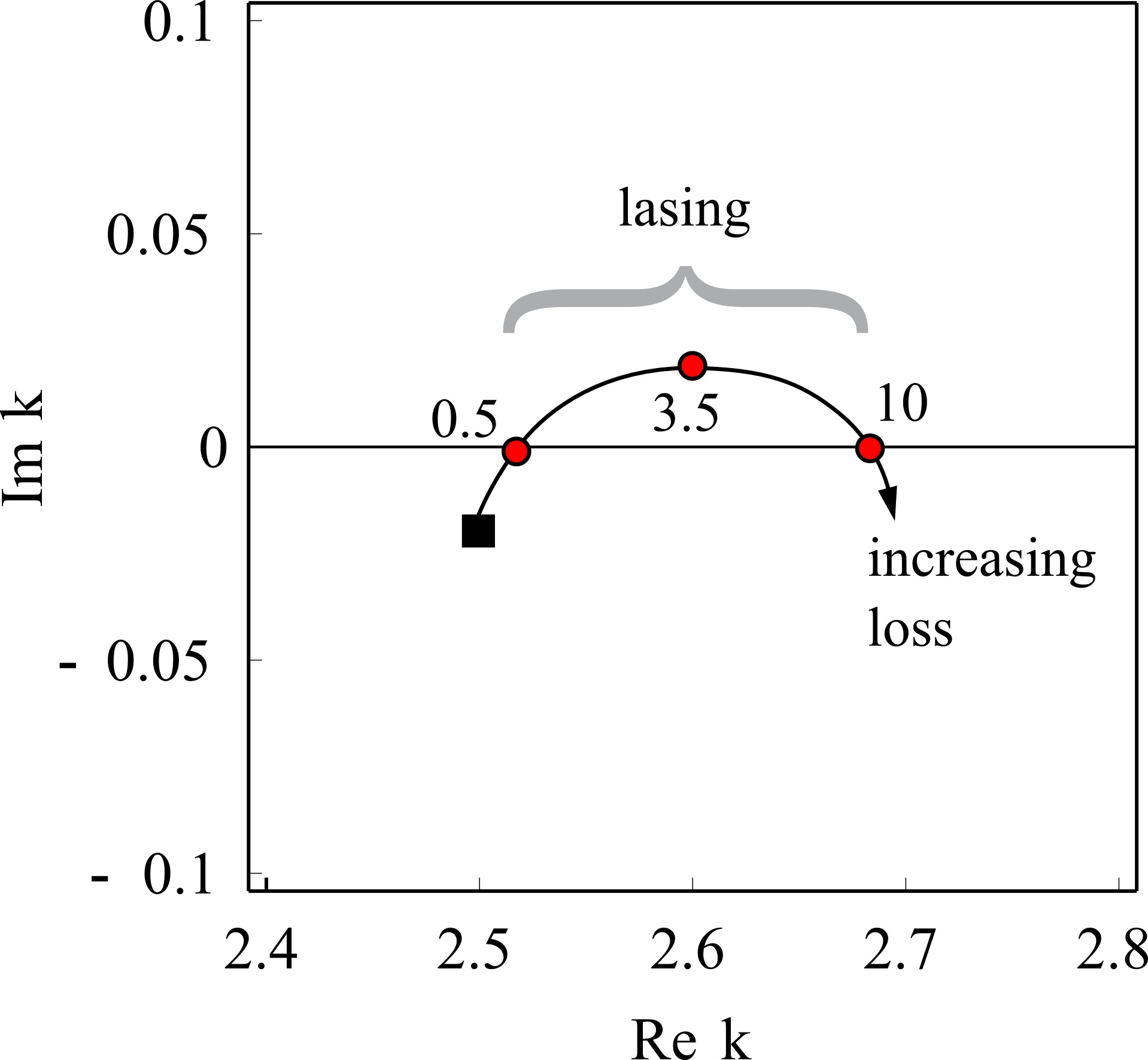}
\caption{\label{fig3} Trajectory of the $S-$matrix pole responsible for absorption induced lasing at fixed gain $f_G=1$. The black box shows the point of poles condensation from which the pole begins to move. Numerical values on the graph are that of the absorption strength $f_A$ at the corresponding point of the pole trajectory.}
\end{figure}

Let us analyze in more detail the trajectrory the $S-$matrix pole demonstrating AIL with increase of loss given that gain is set to the value $f_G=1$. The pole whose trajectory is shown in Fig.~\ref{fig3} emerges at the point of poles condensation located at $k = k_A-i \gamma_A$, as discussed above. At $f_A=1$ it crosses the real axis and continues to move to the upper half-plane. When loss reaches value $f_A=3.5$ the pole begins to move downward towards the real axis. Eventually, with further increase of absorption $f_A$ the pole returns to the lower half-plane. However, before that occurs, another pole associated with the absorbing Lorenzian medium enters the upper half-plane so that laser generation does not stop. This happens until loss exceeds gain in the whole range of frequencies and laser generation shuts down.

Above we demonstrated the phenomenon of AIL for the case of red-shifted absorption frequency. Nevertheless, such behavior of a laser is robust against absorption medium frequency detuning: in simulations we observed occurence of AIL for blue-shited absorption transition and for zero detuning between the emission and absorption ferquency as well. However, in the case of matched emission and absorption frequencies the position of absorption induced pole on the complex frequency plane is such that the lasing instability growth rate is much smaller than the emission linewidth: $\operatorname{Im} \omega \ll \gamma_G$. That means that spontaneous decay of the gain medium happens much faster than the laser generation builds up, so that in practice coherent lasing oscillations would be practically indistinguishable from the broadband spontaneous decay spectrum.

\section{Absorption induced lasing in spherical resonators} 
Toy Fabri-Perot-like cavity analyzed above seems to be not the best candidate for experimental verification of AIL. The above analysis is carried out for normal incidence only, while lasing mode can be, in principle, encountered for oblique incidence at lower values of gain. In fact, there are always poles in the upper half-plane for sufficiently large incidence angle \cite{Nistad}. Having appeared, those oblique modes will clamp population inversion of the gain medium and will strongly affect behavior of normally incident modes. In order to avoid this difficulty, we suggest alternative laser system with spherical symmetry. Being inspired by progress in fabrication of silicon spherical nanoparticles by laser printing \cite{Evlyukhin,Zywietz}, we consider a composite nanoparticle consisting of a dielectric core with gain and lossy shell. Laser modes of such a core-shell nanoparticle are associated with the singularities of Mie coefficients \cite{Liberal, Mie}. Due to the spherical symmetry of this type of laser, its response is isotropic and analysis of lasing modes with repsect to the incidence angle is uneccessary. The choice of non-unifrom distribution of gain and loss in current geometry is reasoned by the simplicity of experimental realization of loss increase on the surface of the spherical resonator. For instance, adhesion of absorbing molecules by the surface will result in an effective absorbing layer.

\begin{figure}  
\includegraphics[width=0.4\textwidth]{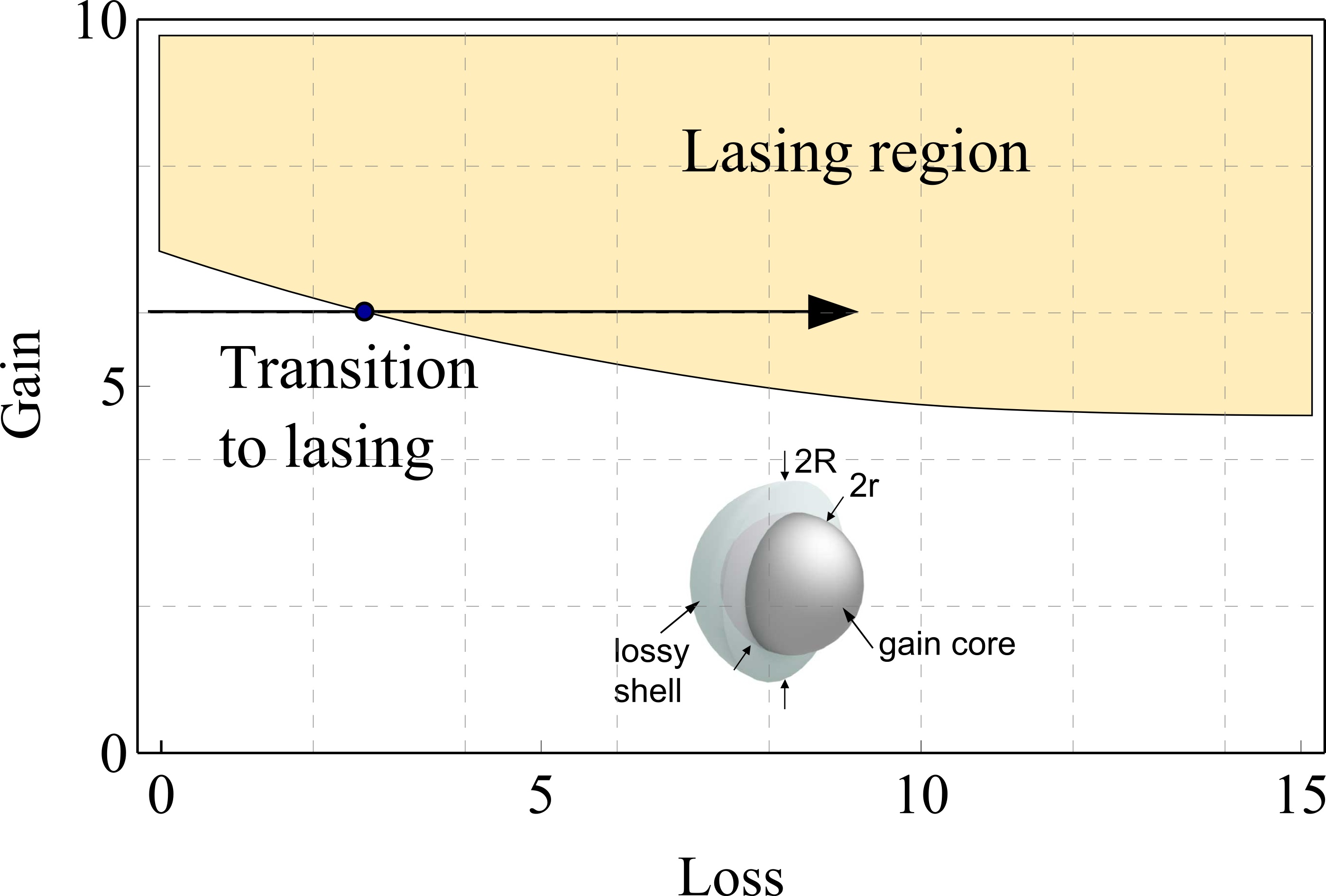}
\caption{\label{fig4}Lasing region in the gain-loss parameter space for the spherical laser resonator. Inset: schematic of the spherical cavity consisting of dielectric core of radius $r= 100 nm$ and shell of thickness $R-r=50 nm$. Background permittvities of core and shell are $\varepsilon_{core}=\varepsilon_{shell}=4$, respectively. Gain is applied to the dielectric core at frequency $k_G=140 nm^{-1}$ and has linewidth $\gamma_G=30 nm^{-1}$. Absorption line is centered at frequency $k_A=142 nm^{-1}$ with linewidth $\gamma_A=0.03\gamma_G$.}
\end{figure}

Up to a constant factor, eigenvalues of the spherical particle $S-$matrix are the scattering Mie coefficients $a_n$ and $b_n$ \cite{Bohren}. In order to analyze behavior of the whole set of  the spherical particle laser modes, we investigate its scattering cross-section defined as ${Q_{sc}} = {{2\pi \sum\limits_{m = 1}^\infty  {\left( {\left( {2m + 1} \right)\left( {{{\left| {{a_m}} \right|}^2} + {{\left| {{b_m}} \right|}^2}} \right)} \right)} } \mathord{\left/
 {\vphantom {{2\pi \sum\limits_{m = 1}^\infty  {\left( {\left( {2m + 1} \right)\left( {{{\left| {{a_m}} \right|}^2} + {{\left| {{b_m}} \right|}^2}} \right)} \right)} } {{k^2}}}} \right.
 \kern-\nulldelimiterspace} {{k^2}}}$. The sum runs over infinite set of indices, however, only finite number of resonances lie within the emission line of gain medium. For the $m-$th resonance of a spherical dielectric particle the very approximate condition determining both electric and magnetic resonance may be written as $k\sqrt \varepsilon  R \sim m$ \cite{Matsko}. Therefore, high order resonances occur at high frequencies far beyond the emission line of gain medium and do not contribute to lasing. The same conclusion can be applied to the core-shell particle with close refractive indeces of core and shell. Fig.~\ref{fig4} shows the generation region of the core-shell nanoresonator in the space of gain and loss parameters. Lasing region is found as the area in the loss-gain parameter space whithin which at least one singularity of the nanoparticle cross-section $Q_{sc}$ lies in the upper half-plane \footnote{When two or more $S-$matrix poles enter the upper half-plane $\operatorname{Im} k > 0$, the laser system may demonstrate either modes competition or complicated multimode regime \cite{Milonni}. Nevertheless, the trivial solution with zero electromagnetic fields becomes unstable and laser oscillations will start.}. Again, introducing of a narrow absorption line leads to emergence of additional poles, which, with increase of absorption strength, enter the lasing region. This observation suggests that phenomenon of AIL is universal and can be realized in numerous laser systems with different geometries.

\section{Conslusion} To conclude, we have shown that an increase of absorption in a laser with uniform spatial distribution of gain and absorbing medium can induce very unusual behavior. We observe that, if the laser is initially in the overpumped state, i.e., such state of a system when gain exceeds loss but the system does not lase, than addition of absorbing medium with narrow linewidth is followed by the transition to lasing. The onset of lasing is due to additional scattering singularities, associated with bulk oscillations of dispersive Lorenzian medium. The transition to lasing is robust against detuning between the gain and absorbing media frequencies. The predicted effect is encountered in planar Fabry-Perot resonator and spherical core-shell dielectric-based nanolaser which can be fabricated by laser printing and doping.

\begin{acknowledgments}
We thank Yu. E. Lozovik for fruitful duscussion. The work was partly supported by RFBR projects No 12-02-01093, 13-02-92660, by the Advanced Research Foundation and by Dynasty Foundation.
\end{acknowledgments}

\bibliography{ail}


%

\end{document}